\documentclass{elsartL}
\usepackage{graphicx}
\usepackage{amssymb}
\newsavebox{\myhbar}
\savebox{\myhbar}{$\hbar$}
\usepackage{amsmath}
\usepackage{mathrsfs}
\usepackage{mathtools}
\usepackage{appendix}
\newcommand{\avg}[1]{\left< #1 \right>} % For a nice display of the average value
\begin{document}

\begin{frontmatter}
\title{Comment on ``A phenomenological $\pi^- p$ scattering length from pionic hydrogen''}
\author{Evangelos Matsinos}
\begin{abstract}
This short technical note addresses a number of issues regarding the estimates of the 2004 paper by Ericson, Loiseau, and Wycech \cite{Ericson2004} for the corrections $\delta_\epsilon$ and $\delta_\Gamma$, which aim at 
the removal of the effects of electromagnetic origin from the measurements of the strong-interaction shift $\epsilon_{1s}$ and of the total decay width $\Gamma_{1s}$ of the ground state in pionic hydrogen.
\end{abstract}
%13.75.Gx: neutron-pion interactions, nucleon-meson interactions, nucleon-pion interactions, pion-baryon reactions, proton-pion interactions
%25.80.Dj: elastic scattering (meson-induced reactions), elastic scattering (pion-nucleus)
%25.80.Gn: charge-exchange reactions (pion), pion absorption and capture
%11.30.-j: symmetry in theory of fields and particles, conservation laws fields and particles
\end{frontmatter}

While occupying myself with the write-up of a review article about the interaction between pions ($\pi$) and nucleons ($N$) at low energy \cite{Matsinos2022a}, I skimmed over (admittedly for the $n$-th time) the 2004 paper 
by Ericson, Loiseau, and Wycech \cite{Ericson2004}. The objective in that work was the determination of corrections to the two scattering lengths $a_{\rm cc}$ and $a_{\rm c0}$, obtained from measurements at the $\pi N$ 
threshold (vanishing laboratory kinetic energy $T$ of the incoming pion), so that the effects of electromagnetic (EM) origin be removed and the purely hadronic scattering lengths $\tilde{a}_{\rm cc}$ and $\tilde{a}_{\rm c0}$ 
be obtained; the quantities $\tilde{a}_{\rm cc}$ and $\tilde{a}_{\rm c0}$ represent the (hadronic) scattering amplitudes of the $\pi^- p$ elastic-scattering (ES) reaction ($\pi^- p \to \pi^- p$) and of the $\pi^- p$ 
charge-exchange (CX) reaction ($\pi^- p \to \pi^0 n$) at $T = 0$ MeV, respectively. The two `untreated' scattering lengths $a_{\rm cc}$ and $a_{\rm c0}$ are extracted from the measurements of the strong-interaction shift 
$\epsilon_{1s}$ and of the total decay width $\Gamma_{1s}$ of the ground state in pionic hydrogen via the so-called ``Deser formulae'' \cite{Deser1954,Trueman1961}, also known by a variety of other names, e.g., as 
Deser-Goldberger-Baumann-Thirring, Deser-Trueman, and Trueman-Deser formulae. Several correction schemes, leading from $a_{\rm cc}$ and $a_{\rm c0}$ to $\tilde{a}_{\rm cc}$ and $\tilde{a}_{\rm c0}$, have been developed, 
one of which has been put forward in Ref.~\cite{Ericson2004}. I decided to pay some further attention to that work, and examine its input in more detail than I had done in the past.

The EM corrections are usually expressed in the form of two quantities, $\delta_\epsilon$ for $a_{\rm cc}$ and $\delta_\Gamma$ for $a_{\rm c0}$. The two hadronic scattering lengths $\tilde{a}_{\rm cc}$ and $\tilde{a}_{\rm c0}$ 
are obtained from $a_{\rm cc}$ and $a_{\rm c0}$ as follows.
\begin{equation} \label{eq:EQ001}
\tilde{a}_{\rm cc} = a_{\rm cc} / (1 + \delta_\epsilon)
\end{equation}
\begin{equation} \label{eq:EQ002}
\tilde{a}_{\rm c0} = a_{\rm c0} / (1 + \delta_\Gamma)
\end{equation}

The study of Ref.~\cite{Ericson2004} was based on the use of Coulomb wavefunctions, modified to account for the effects of the strong interaction, of the extended charge distributions of the interacting hadrons, and of the 
vacuum polarisation. The exertion of the strong interaction is confined within a sphere of radius $R=\avg{r}_{\rm em}$ about the origin ($r=0$) and the electrical charge of the proton is assumed to reside on the surface of 
that sphere, thus yielding constant Coulomb potential in the interior volume. The two solutions for the wavefunction, corresponding to the interior and exterior of that sphere, are matched at $r=R$. This operation perturbs 
the wavefunction at the origin and gives rise to the so-called renormalisation-correction contribution to $\delta_\epsilon$ and $\delta_\Gamma$. The authors obtained analytical expressions for three of the contributions to 
$\delta_\epsilon$: the first term ($\delta^{\avg{r}}$) takes account of the extended charge distribution of the interacting hadrons; the second ($\delta^{\rm c}$) is the renormalisation-correction contribution; the third 
($\delta^{\rm g}$) follows from gauge invariance and ensures that the interaction mimics the scattering off an extended charge Coulomb potential close to the origin. A fourth term ($\delta^{\rm vp}$) is added by hand, to 
account for the effects of the vacuum polarisation. Within the context of the model of Ref.~\cite{Ericson2004}, the correction $\delta_\epsilon$ may be obtained from Eq.~(18) therein as the sum 
$\delta^{\avg{r}}+\delta^{\rm vp}+\delta^{\rm c}+\delta^{\rm g}$, where
\begin{equation} \label{eq:EQ003}
\delta^{\avg{r}}=-2 \frac{\avg{r}_{\rm em}}{r_B} \, \, \, ,
\end{equation}
\begin{equation} \label{eq:EQ004}
\delta^{\rm vp} = 2 \frac{(\delta \psi(0))^{\rm vp}}{\psi(0)} \, \, \, ,
\end{equation}
\begin{equation} \label{eq:EQ005}
\delta^{\rm c} = 2 \hbar c \left( 2 - \gamma - \avg{\ln \frac{2 r}{r_B}}_{\rm em} \right) \frac{\tilde{a}_{\rm cc}}{r_B} \, \, \, ,
\end{equation}
and
\begin{equation} \label{eq:EQ006}
\delta^{\rm g} = 2 \mu \avg{\frac{a}{r}}_{\rm em} \frac{\tilde{b}_{\rm cc}}{\tilde{a}_{\rm cc}} \, \, \, .
\end{equation}
The quantities, appearing in these expressions, are defined as follows.
\begin{itemize}
\item $r_B$ is the Bohr radius $\hbar c/(\mu \alpha)$,
\item $\mu$ is the reduced mass of the $\pi^- p$ system~\footnote{All rest masses and $3$-momenta are expressed in energy units herein. The same goes for the scattering lengths (inverse energy units) and for the range 
parameters (inverse cube energy units).},
\item $\alpha$ is the fine-structure constant,
\item $\psi(r)$ is the non-relativistic Bohr wavefunction of a point charge,
\item $\gamma \approx 0.577216$ is the Euler constant, and
\item $\tilde{b}_{\rm cc}$ is a range parameter for the $\pi^- p$ ES reaction, i.e., the coefficient of the $q^2$ term in the low-energy $s$-wave modelling of the (physical) $\pi^- p$ ES channel~\footnote{The values of 
the isoscalar $b^+$ and isovector $b^-$ range parameters, given in the first paragraph in Section 3 (`Numerical results') of Ref.~\cite{Ericson2004}, are one order of magnitude smaller than the actual values. This is an 
obvious mistype, of no consequence as far as the results of Ref.~\cite{Ericson2004} are concerned.}, see Eq.~(5) of Ref.~\cite{Ericson2004}, where $q$ is the magnitude of the centre-of-mass (CM) $3$-momentum.
\end{itemize}
The vacuum-polarisation correction to the ground-state wavefunction at the origin is given in Table II of Ref.~\cite{Eiras2000} in the form $\alpha^{-1} (\delta \psi(0))^{\rm vp}/\psi(0)$ for several exotic atoms; the 
numerical value for pionic hydrogen is $0.33064$.

The $\delta_\Gamma$ correction was obtained in Ref.~\cite{Ericson2004} as the sum $(\delta^{\avg{r}}+\delta^{\rm vp})/2+\delta^{\rm c}+\delta_0^{\rm g}$, where the first three contributions have already been given in 
Eqs.~(\ref{eq:EQ003}-\ref{eq:EQ005}) and
\begin{equation} \label{eq:EQ007}
\delta_0^{\rm g} = \frac{1}{2} \left( 2 \mu \avg{\frac{a}{r}}_{\rm em} + q_0^2 \right) \frac{\tilde{b}_{\rm c0}}{\tilde{a}_{\rm c0}} \, \, \, ;
\end{equation}
$q_0$ is the magnitude of the CM $3$-momentum of the $\pi^0 n$ system, evaluated at the $\pi^- p$ threshold, and $\tilde{b}_{\rm c0}$ is a range parameter for the $\pi^- p$ CX reaction.

To obtain estimates for the corrections $\delta_\epsilon$ and $\delta_\Gamma$, Ref.~\cite{Ericson2004} used the input values listed in Table \ref{tab:Parameters}, left column of results. In this work, estimates for the 
scattering lengths and range parameters were obtained from fits to the $\pi N$ data for $T \leq 100$ MeV using a simple polynomial parameterisation (along the lines of Ref.~\cite{Ericson2004}) of the spin-isospin $K$-matrix 
elements: $\tilde{K}^{I}_{0+} = \tilde{a}^{I}_{0+} + \tilde{b}^{I}_{0+} q^2 + \tilde{c}^{I}_{0+} q^4$ for the $s$ waves and $\tilde{K}^{I}_{1\pm} = \tilde{a}^{I}_{1\pm} q^2 + \tilde{b}^{I}_{1\pm} q^4$ for the $p$ waves; 
all details can be found in Ref.~\cite{Matsinos2022b}. Explicitly added to the (background) $K$-matrix elements in the $P_{33}$ and $P_{11}$ partial waves were the contributions from the nearby resonances $\Delta(1232)$ and 
$N(1440)$, respectively. The results of this work are listed in the right column of the table.

The two sets of input values yield the corrections of Table \ref{tab:Corrections}. The differences in the gauge-term contributions are substantial and emanate from the use of outdated values for the range parameters 
$\tilde{b}_{\rm cc}$ and $\tilde{b}_{\rm c0}$ in Ref.~\cite{Ericson2004}, imported therein from the Karlsruhe analyses \cite{Hoehler1983}. The updated $\delta_\Gamma$ correction (see Appendix \ref{App:AppA}) matches the 
result of Ref.~\cite{Zemp2004}, obtained within the framework of Chiral-Perturbation Theory also in 2004, though this agreement is, most likely, coincidental given the disparate approaches employed in the two studies.

I need to emphasise (once again) that the results of the Karlsruhe analyses had been based on fewer than one hundred low-energy $\pi N$ measurements in total, almost exclusively of the $\pi^+ p$ differential cross section 
(DCS). No measurements of the analysing power (which lead to an improved determination of the three small $p$-wave amplitudes) were available; nor were any partial-total cross sections at that time. Most of the then-available 
$\pi^+ p$ DCS measurements turned out to be incompatible (due to the shape of the angular distribution of their DCS and/or to their absolute normalisation) with the modern measurements, obtained later at the three meson 
factories: LAMPF, PSI, and TRIUMF. The extraction in the Karlsruhe analyses of the $\pi N$ partial-wave amplitudes at low energy rested upon overconstrained fits to the data available at higher energies.

My sole aim in this note was to provide updated results for the corrections $\delta_\epsilon$ and $\delta_\Gamma$ of Ref.~\cite{Ericson2004}, and raise awareness of the perils of importing the outdated values of the 
Karlsruhe programme into modern studies. This work was not intended as an assessment of the correctness/applicability/integrity of the approach put forward in Ref.~\cite{Ericson2004}, nor should it be taken as either 
corroborating or refuting that study. The reader ought to bear in mind that criticism about that approach has already been expressed, e.g., see Section 4 of Ref.~\cite{Oades2007}. On the other hand, Ericson, Loiseau, and 
Wycech have also expressed their concerns about the incompatibility of the coupled-channel formalism with the low-energy expansion of the $\pi N$ $K$-matrix elements \cite{Ericson2004}.

\begin{ack}
I would like to thank T.E.O.~Ericson for answering a few questions regarding Ref.~\cite{Ericson2004}. Any misinterpretation of the material of Ref.~\cite{Ericson2004} should be blamed on me.
\end{ack}

\clearpage
\vspace{0.5cm}
\begin{table}%[h!]
{\bf \caption{\label{tab:Parameters}}}The values of the quantities needed for the evaluation of the corrections $\delta_\epsilon$ and $\delta_\Gamma$ in Ref.~\cite{Ericson2004}. For the sake of conformity, all entries are 
expressed either in fm or in (powers of) GeV. The first column of values corresponds to the input used in Ref.~\cite{Ericson2004}, the second to the input in this work. The ratios between each range parameter and the 
corresponding scattering length (i.e., $\tilde{b}_{\rm cc}/\tilde{a}_{\rm cc}$ and $\tilde{b}_{\rm c0}/\tilde{a}_{\rm c0}$ ) had been evaluated in Ref.~\cite{Ericson2004} assuming independent determinations of these 
quantities. The same ratios are evaluated in this work from the outcome of the fits to the low-energy $\pi^- p$ ES and CX databases, and also take into account the (strong) \emph{anti}correlation between the quantities 
involved in these ratios \cite{Matsinos2022b}. All other physical constants (i.e., rest masses, conversion constants, etc.) have been fixed from the 2020 compilation of the Particle-Data Group \cite{PDG2020}.
\vspace{0.25cm}
\begin{center}
\begin{tabular}{|l|c|c|}
\hline
Quantity & Ref.~\cite{Ericson2004} & This work \\
\hline
\hline
$\avg{r}_{\rm em}$ (fm) & $0.95(1)$ \cite{Ericson1988} & Unchanged \\
$\avg{\alpha/r}_{\rm em}$ (GeV) & $2.13(2) \cdot 10^{-3}$ \cite{Ericson1988} & Unchanged \\
$\avg{\ln (\mu r/(\hbar c))}_{\rm em}$ & $-0.687(9)$ \cite{Ericson1988} & Unchanged \\
$\tilde{a}_{\rm cc}$ (GeV$^{-1}$) & $0.6233(36)$ \cite{Ericson2004} & $0.6129(33)$ \\
$\tilde{b}_{\rm cc}$ (GeV$^{-3}$) & $-11.4(3.3)$ \cite{Hoehler1983} & $-18.2(1.0)$ \\
$\tilde{b}_{\rm cc}/\tilde{a}_{\rm cc}$ (GeV$^{-2}$) & $-18.3(5.3)$ & $-29.7(1.6)$ \\
$\tilde{a}_{\rm c0}$ (GeV$^{-1}$) & $-0.896(29)$ \cite{Ericson2004} & $-0.925(11)$ \\
$\tilde{b}_{\rm c0}$ (GeV$^{-3}$) & $-7.0(3.3)$ \cite{Hoehler1983} & $1.7(2.1)$ \\
$\tilde{b}_{\rm c0}/\tilde{a}_{\rm c0}$ (GeV$^{-2}$) & $7.8(3.7)$ & $-1.8(2.3)$ \\
\hline
\end{tabular}
\end{center}
\vspace{0.5cm}
\end{table}

\vspace{0.5cm}
\begin{table}%[h!]
{\bf \caption{\label{tab:Corrections}}}The corrections $\delta_\epsilon$ and $\delta_\Gamma$, obtained in Ref.~\cite{Ericson2004} and in this work. The differences are almost entirely due to the use of outdated values for 
the range parameters $\tilde{b}_{\rm cc}$ and $\tilde{b}_{\rm c0}$ in Ref.~\cite{Ericson2004}, imported therein from the Karlsruhe analyses \cite{Hoehler1983}. The values of the quantities, needed for the evaluation of 
these corrections, were given in Table \ref{tab:Parameters}.
\vspace{0.25cm}
\begin{center}
\begin{tabular}{|l|c|c|}
\hline
Contribution & Ref.~\cite{Ericson2004} & This work \\
\hline
\hline
\multicolumn{3}{|c|}{$\delta_\epsilon$ (\%)}\\
\hline
Finite size & $-0.8537(90)$ & Unchanged \\
Vacuum polarisation & $0.4826$ & Unchanged \\
Renormalisation & $0.7004(41)$ & $0.6887(38)$ \\
Gauge term & $-0.95(28)$ & $-1.537(85)$ \\
\hline
Total & $-0.62(28)$ & $-1.219(85)$ \\
\hline
\hline
\multicolumn{3}{|c|}{$\delta_\Gamma$ (\%)}\\
\hline
Finite size & $-0.4268(45)$ & Unchanged \\
Vacuum polarisation & $0.2413$ & Unchanged \\
Renormalisation & $0.7004(41)$ & $0.6887(38)$ \\
Gauge term & $0.51(24)$ & $-0.12(15)$ \\
\hline
Total & $1.03(24)$ & $0.39(15)$ \\
\hline
\end{tabular}
\end{center}
\vspace{0.5cm}
\end{table}

\clearpage
\newpage
\appendix
\section{\label{App:AppA}Extraction of $\tilde{b}_{\rm cc}$ and $\tilde{b}_{\rm c0}$ from the modern data}

Although a `short technical note' is hardly expected to contain any appendices, I found no better way to deal with the extraction of meaningful estimates for the range parameters $\tilde{b}_{\rm cc}$ and $\tilde{b}_{\rm c0}$, 
the way they were defined in Ref.~\cite{Ericson2004}. The $s$-wave $K$-matrix element of the (physical) $\pi^- p$ ES channel is expanded therein as
\begin{equation} \label{eq:EQA001}
\tilde{K}_{\rm cc} = \tilde{a}_{\rm cc} + \tilde{b}_{\rm cc} q^2 + \mathcal{O}(q^4) \, \, \, .
\end{equation}
In their single-channel formalism, the relation between the $T$-matrix element $\tilde{T}_{\rm cc}$ and $\tilde{K}_{\rm cc}$ should read as
\begin{equation} \label{eq:EQA002}
\tilde{T}_{\rm cc} = \frac{\tilde{K}_{\rm cc}}{1 - i q \tilde{K}_{\rm cc}} \Rightarrow \Re [\tilde{T}_{\rm cc}] = \frac{\tilde{K}_{\rm cc}}{1 + q^2 \tilde{K}^2_{\rm cc}} = \tilde{K}_{\rm cc} \left( 1 - q^2 \tilde{K}^2_{\rm cc} + \mathcal{O}(q^4) \right) \, \, \, .
\end{equation}
Replacing $\tilde{K}_{cc}$ from Eq.~(\ref{eq:EQA001}), one obtains
\begin{equation} \label{eq:EQA003}
\Re [ \tilde{T}_{\rm cc} ] = \tilde{a}_{\rm cc} + \left( \tilde{b}_{\rm cc} - \left( \tilde{a}_{\rm cc} \right)^3 \right) q^2 + \mathcal{O}(q^4) \, \, \, .
\end{equation}
The relation between each partial-wave amplitude $\tilde{T}^I_{l\pm}$ and the corresponding $K$-matrix element $\tilde{K}^I_{l\pm}$ reads as
\begin{equation} \label{eq:EQA004}
\tilde{T}^I_{l\pm} = \frac{\tilde{K}^I_{l\pm}}{1-i q \tilde{K}^I_{l\pm}} \Rightarrow \Re [\tilde{T}^I_{l\pm}] = \tilde{K}^I_{l\pm} \left( 1 - q^2 \left( \tilde{K}^I_{l\pm} \right)^2 + \mathcal{O}(q^4) \right) \, \, \, ,
\end{equation}
where the total isospin $I$ is either $3/2$ or $1/2$, and the subscripts denote the total angular momentum $j=l \pm 1/2$. As the interest is in the $s$ waves, $\tilde{T}_{\rm cc} = (\tilde{T}^{3/2}_{0+} + 2 \tilde{T}^{1/2}_{0+})/3$ and
\begin{equation} \label{eq:EQA005}
\Re [\tilde{T}_{\rm cc}] = \frac{1}{3} \left( \tilde{K}^{3/2}_{0+} + 2 \tilde{K}^{1/2}_{0+} - \left( \left( \tilde{K}^{3/2}_{0+} \right)^3 + 2 \left( \tilde{K}^{1/2}_{0+} \right)^3 \right) q^2 + \mathcal{O}(q^4) \right) \, \, \, .
\end{equation}
Retaining the terms up to $\mathcal{O}(q^2)$, the usual low-energy expansion of the spin-isospin $K$-matrix elements in the $s$ waves reads as
\begin{equation} \label{eq:EQA006}
\tilde{K}^{3/2}_{0+} \approx \tilde{a}^{3/2}_{0+} + \tilde{b}^{3/2}_{0+} q^2 \, \, \, \text{and} \, \, \, \tilde{K}^{1/2}_{0+} \approx \tilde{a}^{1/2}_{0+} + \tilde{b}^{1/2}_{0+} q^2 \, \, \, .
\end{equation}
From Eqs.~(\ref{eq:EQA003},\ref{eq:EQA005},\ref{eq:EQA006}), one obtains
\begin{equation} \label{eq:EQA007}
\tilde{b}_{\rm cc} = \frac{\tilde{b}^{3/2}_{0+} + 2 \tilde{b}^{1/2}_{0+}}{3} + \left( \tilde{a}_{\rm cc} \right)^3 - \frac{ \left( \tilde{a}^{3/2}_{0+} \right)^3 + 2 \left( \tilde{a}^{1/2}_{0+} \right)^3}{3} \coloneqq \frac{\tilde{b}^{3/2}_{0+} + 
2 \tilde{b}^{1/2}_{0+}}{3} + \Delta \tilde{b}_{\rm cc} \, \, \, ,
\end{equation}
where the quantities, which must be raised to a power, are placed in parentheses; of course,
\begin{equation} \label{eq:EQA007_5}
\tilde{a}_{\rm cc} = \frac{\tilde{a}^{3/2}_{0+} + 2 \tilde{a}^{1/2}_{0+}}{3} \, \, \, .
\end{equation}
The $\Delta \tilde{b}_{\rm cc}$ contribution to the $\tilde{b}_{\rm cc}$ value of Table \ref{tab:Parameters} is equal to $-0.764(41)$ GeV$^{-3}$.

Similarly,
\begin{equation} \label{eq:EQA008}
\tilde{b}_{\rm c0} = \sqrt{2} \, \frac{\tilde{b}^{3/2}_{0+} - \tilde{b}^{1/2}_{0+}}{3} + \left( \tilde{a}_{\rm c0} \right)^3 - \sqrt{2} \, \frac{ \left( \tilde{a}^{3/2}_{0+} \right)^3 - \left( \tilde{a}^{1/2}_{0+} \right)^3}{3} \coloneqq 
\sqrt{2} \, \frac{\tilde{b}^{3/2}_{0+} - \tilde{b}^{1/2}_{0+}}{3} + \Delta \tilde{b}_{\rm c0} \, \, \, ,
\end{equation}
where
\begin{equation} \label{eq:EQA008_5}
\tilde{a}_{\rm c0} = \sqrt{2} \, \frac{\tilde{a}^{3/2}_{0+} - \tilde{a}^{1/2}_{0+}}{3} \, \, \, .
\end{equation}
The $\Delta \tilde{b}_{\rm c0}$ contribution to the $\tilde{b}_{\rm c0}$ value of Table \ref{tab:Parameters} is equal to $0.84(12)$ GeV$^{-3}$.

\end{document}